\documentclass[aps,prl,preprint,showpacs,superscriptaddress]{revtex4}
\usepackage{graphicx}
\preprint{Draft}
\newcommand{\beq}{\begin{eqnarray}}
\newcommand{\eeq}{\end{eqnarray}}

\begin{document}
\title{Temperature-dependent properties of the magnetic order in single-crystal BiFeO$_3$}

\author{M. Ramazanoglu}
\affiliation{Department of Physics and Astronomy, Rutgers University,
Piscataway, NJ 08854, USA}
\author{W. Ratcliff II}
\affiliation{NIST Center for Neutron Research, National Institute of
Standards and Technology, Gaithersburg, MD 20899, USA}
\author{Y. J. Choi}
\affiliation{Department of Physics and Astronomy, Rutgers University,
Piscataway, NJ 08854, USA}
\author{Seongsu Lee}
\affiliation{Neutron Science Division, Korea Atomic Research
Institute, Daejon 305-353, Korea}
\author{S.-W. Cheong}
\affiliation{Department of Physics and Astronomy, Rutgers University,
Piscataway, NJ 08854, USA}
\author{V. Kiryukhin}
\affiliation{Department of Physics and Astronomy, Rutgers University,
Piscataway, NJ 08854, USA}

\date{\today}
\begin{abstract}

We report neutron diffraction and magnetization studies of the magnetic
order in multiferroic BiFeO$_3$. In ferroelectric monodomain single
crystals, there are three magnetic cycloidal domains with propagation
vectors equivalent by crystallographic symmetry. The cycloid period slowly
grows with increasing temperature. The magnetic domain populations do not
change with temperature except in the close vicinity of the N\'eel
temperature, at which, in addition, a small jump in magnetization is
observed. No evidence for the spin-reorientation transitions proposed in
previous Raman and dielectric studies is found. The magnetic cycloid is
slightly anharmonic for $T$=5 K. The anharmonicity is much smaller than
previously reported in NMR studies. At room temperature, a circular
cycloid is observed, within errors. We argue that the observed
anharmonicity provides important clues for understanding electromagnons in
BiFeO$_{3}$.

\end{abstract}
\maketitle
\section{Introduction}

The coupling between ferroelectric (FE) polarization and magnetic order in
solid state materials, known as multiferroics, attracts significant
interest because of its importance for understanding the properties of
these materials, as well as of engineering application
possibilities\cite{Catalan, Cheong, Ramesh, Zvezdin1}. Among the known
multiferroics, BiFeO$_{3}$ (BFO) is the only one exhibiting both magnetic
and FE orders at room temperature. These orders are coupled, and control
of both FE and magnetic properties by an electric field has been
demonstrated in thin film and single crystal samples,\cite{Ramesh2,
Valery1} as well as discussed theoretically.\cite{Ederer,Lisenkov}
Importantly, the magnitude of the spontaneous electric polarization
($P\sim 10^2 \mu C/cm^2$) in BFO is very large, facilitating control by
the electric field, and opening possibilities for applications in
multistate memory devices, spintronics, and configurable
electronics.\cite{Science} Therefore, BFO attracts considerable attention
compared to other multiferroics. Until recently, BFO research was limited
to thin films showing rather different properties from those of the bulk
samples, and to polycrystals.\cite{Catalan} Since suitable single crystals
have become available recently, their properties were subject of many
studies\cite{Valery1,Science,Valery2,Lebeugle,Cazayous,Rovillain,Rovillain2}.
Some of the basic properties of BFO, however, still remain to be
characterized.

The temperature-dependent evolution of the magnetic structure in BFO is an
important subject with several currently unresolved issues. Below the FE
transition at $T_c\sim$1150 K, BFO exhibits an $R3c$ rhombohedral
perovskite structure, which is described in this paper using a
pseudo-cubic notation with $a\sim$3.96\ {\AA}, and
$\alpha\sim$89.4$^\circ$. Fe$^{3+}$ spins order at $T_N\sim$640 K. It has
long been known that the magnetic order is of the antiferromagnetic G
type, with a long-range cycloidal modulation superimposed.\cite{Sosnowska}
This modulation can propagate along three directions equivalent by
symmetry, described by propagation vectors $\tau_1=\delta$(1,-1,0),
$\tau_2=\delta$(1,0,-1), and $\tau_3=\delta$(0,-1,1), where
$\delta\sim$0.0045 r.l.u. For each of the three cycloids, the spins rotate
in the plane defined by the (111) and $\tau$ vectors, see Fig. 1(a,b). A
possibility for a so far undetected canting of the spins out of the
cycloid plane has been discussed theoretically.\cite{Ederer} Below the
room temperature, several anomalies have been observed in
Raman\cite{Cazayous,Singh1} and dielectric susceptibility
measurements.\cite{Redfern} Some of these anomalies, such as those at
$T$=140 K and 200 K, are quite significant. It was proposed that these
anomalies were associated with putative spin reorientation transitions,
which are rather common in magnetic orthoferrites.\cite{Catalan} We note
that spin reorientation can occur through magnetic transitions within the
existing magnetic domains, or through changes in the fractional
populations of the equivalent magnetic domains $\tau_i$. NMR measurements
indicate that the spin cycloid is anharmonic at $T$=5 K, while the
anharmonicity is absent at $T$=300 K.\cite{Zalesskii,Zalesskii2,Bush}
However, this effect is subtle, and therefore unlikely to produce the
strong anomalies at $T$=140 K and 200 K. In any case, the NMR measurements
of the magnetic order are indirect, and require confirmation. In fact,
neutron diffraction studies of polycrystals did not find any measurable
differences between the magnetic structures at $T$=4 K and at the room
temperature.\cite{Przenioslo} In particular, no dependence of $\delta$ on
temperature was reported. So far, measurements of the magnetic order
parameter and search for possible changes of the magnetic structure by
direct techniques in the full temperature range, from helium temperatures
up to $T_N$, were absent, in part due to absence of suitable single
crystals. Such measurements are clearly necessary to address the questions
discussed above. They are also needed to understand the character of the
magnetic transition at $T_N$ and its possible interrelationship with the
FE properties of BFO. The neutron diffraction and magnetization
measurements described in the current study were motivated by these needs.

\section {Experimental details}

Single crystals of BiFeO$_3$ were grown using flux method as described in
Refs. \cite{Science,Valery2}. The crystal growth occurs below the FE
transition temperature and tends to produce FE monodomain crystals.
Neutron diffraction experiments were carried out on BT-9 triple-axis
spectrometer at NIST Center for Neutron Research. A closed cycle cryostat
and a furnace were used for temperature-dependent measurements. The
neutron energy was 14.7 meV, and collimations of 40-40-S-40-open (order
parameter measurements), or 10-10-S-10-open (third harmonic studies) were
used. To reduce the $\lambda$/2 contamination of the neutron beam,
pyrolytic graphite filters were used before and after the sample, as well
as in-pile. A 3$\times$2$\times$0.5 mm$^3$ FE monodomain sample was
oriented in the ($h$, $k$, $(h+k)/2$) scattering plane. In this paper, we
use the pseudo-cubic notation for all the reciprocal lattice vectors,
which are measured in reciprocal lattice units, r.l.u. Magnetization
measurements were performed using a SQUID magnetometer under an applied
magnetic field of 0.2 Tesla. In all figures, error bars are from counting
statistics.

\section {Results and Discussion}

To characterize temperature-dependent behavior of the magnetic order
parameter, periodicity of the spin cycloid, equivalent magnetic domain
populations, and possible anharmonicity of the cycloid, single-crystal
neutron diffraction experiments were carried out. The most intense
magnetic reflection is observed at the $Q_0\pm \tau_i$ positions, where
$Q_0$=(0.5, 0.5, 0.5), and $i$=1,2,3, giving rise to six diffraction
peaks. The experimental scattering geometry is illustrated in Fig. 1
(c,d). The horizontal scattering plane contains perpendicular vectors
$Q_0$ and $\tau_1$. Therefore, magnetic domain $\tau_1$ (the in-plane
domain) produces peaks at $Q_0\pm\delta$ in the scan along the $\tau_1$
centered at $Q_0$. The magnetic domains $\tau_2$ and $\tau_3$ (the
out-of-plane domains) produce peaks in the plane normal to the scattering
plane. Nevertheless, they are also captured by this scan because of the
broad instrumental resolution normal to the scattering plane. As
illustrated in Fig. 1(d), the out-of-plane domains produce overlapping
peaks at $Q_0\pm\delta/2$. Thus, a single scan along $\tau_1$ at the (0.5,
0.5, 0.5) position allows to characterize the in-plane domain, as well as
to measure the combined signal of the out-of-plane domains.

The scans were taken in the temperature range from $T$=5 K to 650 K. The
scattering profiles are shown in Fig. 2 for several representative
temperatures. For $T>T_N\sim$640 K, a weak temperature-independent peak is
observed at $Q_0$. It can originate from a $\lambda/2$ leakage in the
neutron beam, or reflect a previously undetected small lattice distortion
at this wave vector. In the magnetic phase, two asymmetric peaks centered
roughly at $Q_0\pm\delta$ are observed. These peaks have clear shoulders
in the direction of $Q_0$, as best seen for the $T$=5 K data in Fig. 2.
The overall peak shape is described perfectly by combined signals from the
in- and out-of-plane magnetic domains at $Q_0\pm\delta$, and
$Q_0\pm\delta/2$, respectively. This picture was confirmed by the
room-temperature scans in the planes containing $Q_0$ and either $\tau_2$
or $\tau_3$ (not shown). Therefore, the scattering profiles were fitted
using five Gaussian peaks: at $Q_0\pm\delta$ (the in-plane magnetic
domain), $Q_0\pm\delta/2$ (combined out-of-plane domains), and a weak
temperature-independent background peak at $Q_0$. The common Gaussian peak
width was determined as a global parameter fitted to a common value over
the entire temperature range. The intensities of the magnetic peaks, as
well as the wave vector of the cycloid $\delta$, were fitted at each
temperature. This procedure resulted in high-quality fits, as shown in
Fig. 2.

Figure 3(a) shows the temperature dependence of the integrated magnetic
intensity, which is proportional to the square of the magnetic order
parameter. It exhibits a canonical second-order-like behavior, and is well
described by the power law (1-$T/T_N$)$^{2\beta}$ in a rather large region
below $T_N$, see Fig. 3(a). The power law gives $T_N$=639.6(5) K, and
$\beta$=0.34(3). The critical exponent $\beta$ is in agreement with
three-dimensional magnetic transition. A similar $\beta$ value was
previously found in M\"ossbauer studies.\cite{Blaauw} The transition
temperature T$_N$ is within the range of previously reported values for
polycrystalline samples (590-650 K),\cite{Catalan,Fisher} and is
consistent with theoretical predictions.\cite{Albrecht} The ratio of the
integrated intensity of the magnetic peaks at $Q_0\pm\delta$ to that at
$Q_0\pm\delta/2$ gives the ratio of the in-plane and the out-of-plane
domain populations. This ratio is shown in Fig. 3(b) as a function of
temperature. Except for the narrow vicinity of the magnetic transition
temperature, this ratio is essentially constant. As $T_N$ is approached,
the ratio decreases from about 1.5 to 0.8, {\it i.e.} the relative
population of the out-of-plane domains grows. This observation may reflect
increased significance of thermal fluctuations for $T\sim T_N$. The
magnitude of the cycloidal modulation wave vector $\delta$ decreases with
temperature, as shown in Fig. 3(c). The corresponding modulation period
grows from 629(5) \AA\ at $T$=5 K to 780(30) \AA\ at $T$=615 K. There is
an apparent change of slope of this temperature dependence at $T\sim$400
K. In previous works,\cite{Przenioslo} the measurements were done in a
smaller temperature range, and no temperature dependence of $\delta$ was
reported.

Figure 3(d) shows the temperature dependence of the magnetization, with
the magnetic field applied along different crystallographic directions.
The general behavior below $T_N$ is consistent with an antiferromagnet
containing a small ferromagnetic impurity fraction. Similar behavior was
reported elsewhere.\cite{Catalan,Singh2} The anomaly at $T_N$ itself is,
however, rather unusual. The abrupt upward magnetization jump which
accompanies appearance of the magnetic order is seemingly inconsistent
both with the otherwise second-order-like behavior of the magnetic order
parameter, and with the antiferromagnetic transition. The magnetization
jump is rather small, corresponding to effective magnetic moment of $\sim
5\times 10^{-4}\mu_B$/spin in the applied field of 0.2 Tesla. This
suggests that the spin component involved in the abrupt transition is
small, which is consistent with the order parameter behavior of Fig. 3(a).
Interestingly, the distribution of the magnetic domain populations also
exhibits an abrupt change in the same temperature region. These data
suggest that the magnetic structure of BFO is more complex than currently
believed, or that nontrivial effects related to the magnetic domains occur
near the magnetic transition temperature. The symmetry of the magnetic
structure is a key factor defining possible magnetoelectric effects, such
as existence of linear magnetoelectric coupling.\cite{Kadomtseva} Further
study of the nature of the transition at $T_N$ is, therefore, of
significant interest.

Deviations of the magnetic structure from the circular cycloid have been
detected previously, but only at cryogenic temperatures. Specifically,
line shapes observed in NMR spin-echo measurements at $T$=5 K are
inconsistent with this structure.\cite{Zalesskii,Zalesskii2,Bush} The NMR
results were explained using an anharmonic cycloidal spin structure, which
results from magnetic anisotropy in the plane of the spin rotation. Free
energy density in BFO can be written\cite{Kadomtseva} in the form
$f=f_L+f_{exch}+f_{an}$. Here $f_L$ is the relativistic inhomogeneous
magnetoelectric interaction, $f_{exch}=A\sum (\nabla l_i)^2$ ($i=x,y,z$)
is the inhomogeneous exchange energy with exchange stiffness $A$,
$f_{an}=K_u sin^2(\theta)$ is the anisotropy energy with the anisotropy
constant $K_u$, {\it\bf l} is the antiferromagnetic vector, and $\theta$
is the angle between {\it\bf l} and the (111) direction. The solution of
this model is an anharmonic cycloid in which the spin component along the
(111) direction is given by the elliptic Jacobi function
$sn(4K(m)x/\lambda,m)$, where $m$ is its parameter, $K(m)$ is the complete
elliptic integral of the first kind, $\lambda$ is the modulation period,
and $x$ designates the propagation direction.\cite{Kadomtseva} Depending
on the parameter $m$, the spin modulation changes from purely circular
($m$=0), to a square wave in which the spins bunch along the (111)
direction ($m$=1). NMR line shape analysis based on this model
indicated that the spin cycloid is strongly distorted at $T$=5 K
($m$=0.95), while it becomes essentially circular at room temperature
($m$=0.48).\cite{Zalesskii,Zalesskii2,Bush}

Magnetic neutron diffraction signal is produced by the spin components
normal to the scattering vector. In our experiment, the scattering vector
is essentially parallel to (111), and therefore only the spin components
along the cycloid propagation vector $\tau$ contribute to scattering.
These components are given by
(1-$sn{^2}(4K(m)x/\lambda,m)$)$^{0.5}$=$cn(4K(m)x/\lambda,m)$. The
diffraction intensity is proportional to the square of the Fourier
transform of the magnetization density. The Jacobi function has only odd
harmonics in its Fourier series, and therefore scattering from the
anharmonic cycloid described above should produce peaks at $Q_0\pm n\tau$
($n$ is odd), whose intensity is proportional to the square of the
appropriate Fourier coefficient. The ratio of these coefficients depends
only on $m$. Thus, the ratio of the integrated intensities of the first
($n$=1) to the third ($n$=3) harmonics of the magnetic peak, $I_1/I_3$,
provides direct measurement of $m$. Fig. 4(a) shows high-resolution scans
capturing the $Q_0-\tau_1$ and $Q_0-3\tau_1$ positions, taken at $T$=5 K
and 300 K. A barely discernible third harmonic appears to be present at
$T$=5 K, while none is seen at $T$=300 K. The data in Fig. 4 (a) were
fitted using the model described above with the third harmonic peak added,
and the line shapes changed to the resolution-corrected pseudo-Voigt
function to correctly describe the wings of the peaks. For $T$=5 K, the
fit gives $I_1/I_3$=500$\pm$100 corresponding to $m$=0.50, and for $T$=300
K $I_1/I_3>$2000. The obtained $m$=0.50 produces only a very small
deviation from the circular cycloid structure, as illustrated in Fig.
4(b), and in the inset in Fig. 4(a). In contrast, the value of $m$=0.95
derived from the NMR data at $T$=5 K would imply a clear bunching of spins
along the (111) direction, see Fig. 4(c). We note, however, that $m$=0.95
would give $I_1/I_3\sim$25, which is clearly inconsistent with our
results. Neutron diffraction provides a more direct method for measuring
the anharmonicity than NMR, and therefore we conclude that the
low-temperature cycloid is more harmonic than previously reported. Both of
these techniques indicate that the cycloid is essentially circular at room
temperature.

The anisotropy energy $K_u$ is related\cite{Zalesskii2} to $m$ by
$K_u=16mAK^2(m)/\lambda^2$. The literature values of the exchange
stiffness $A$ are in the range of (2-4)$\times$10$^{-14}$ J/cm.
\cite{Zalesskii, Zalesskii2, Zalessky, Zhdanov, Logginov,
Sosnowska2} Using $m$=0.50, one obtains $K_u$=(1.5-3)$\times$10$^{-2}$
J/cm$^3$=(0.6-1.2)$\times$10$^{-5}$ eV/spin. This is about four times
smaller than that obtained using the NMR value of $m$, and two times
smaller than the value found using ESR.\cite{Ruette} We note that the
magnetic anisotropy is comparable to the inhomogeneous exchange energy
$E_{exch}\sim A\delta^2$ since $K_u/(A\delta^2)=4mK^2(m)/\pi^2=$0.7 for
$m$=0.50. Both of these energies are smaller than $kT$, and therefore the
anharmonic cycloid can be significantly affected by thermal fluctuations.
This could explain, in particular, the apparent disappearance of the third
harmonic peak at high temperatures, since the fluctuation-induced peak
broadening often increases with the harmonic order,\cite{Aharony,RJC} and
any broadening of the already small peak would render it immeasurable.

Anharmonicity of the spin cycloid is an important parameter defining
dynamical magnetoelectric response. It determines which spin wave branches
couple to the electric polarization, giving rise to hybrid excitations
(electromagnons). It was predicted, for instance, that the cyclone magnon
mode $\phi_2$, and the extra-cyclone magnon mode $\psi_3$ only have the
hybrid character in the presence of the third harmonic in the
cycloid.\cite{Sousa} Higher-order harmonics make other cycloidal spin
modes hybrid. Such spin modes have been observed in BFO.\cite{Cazayous}
Importantly, they have been found to exhibit giant energy shifts in an
applied electric field, indicating a significant magnetoelectric
coupling.\cite{Rovillain2} Our measurements of the cycloid anharmonicity
and of the relevant magnetic interaction energies should provide valuable
input for microscopic description of the excitation modes and of the
observed giant magnetoelectric effects in BFO.

As discussed above, various anomalies have been observed below room
temperature in Raman and dielectric property studies. Some of these
anomalies were attributed to putative magnetic transitions. We do not
observe any changes in the magnetic structure for $T<$300 K, except for a
slow and gradual change of the cycloid periodicity. Other recent neutron
data are in agreement with these results.\cite{Herrero-Albillos} Our data
therefore show that the reported anomalies are not related to any magnetic
transition, with only possible exception being a sudden change in the
magnetic domain populations at these temperatures. While the
temperature-dependent behavior of the domain populations could certainly
be sample dependent, the observed absence of such a change in our samples
makes the latter scenario appear unlikely. Other physical mechanisms
should therefore be sought to explain the reported anomalies.

\section {Conclusions}

To summarize, we report neutron diffraction and magnetization studies of
the magnetic properties of single-crystal BiFeO$_3$ in the temperature
range from $T$=5 K to the temperatures exceeding $T_N$. The magnetic order
parameter exhibits a smooth second-order-like dependence typical of the
three-dimensional magnetic transition. On the other hand, the
magnetization exhibits a small jump at $T_N$, which is seemingly
inconsistent with a regular antiferromagnetic transition. The reason for
this behavior is yet to be determined. The period of the magnetic
modulation smoothly increases with temperature from 630 \AA\ at $T$=5 K to
$\sim$780 \AA\ near $T_N$, exhibiting a change of slope at $T\sim$400 K.
The populations of the magnetic domains do not change with temperature,
except in the close vicinity of $T_N$. There is a very small anharmonicity
in the magnetic cycloid at $T$=5 K (the Jacobi function parameter
$m$=0.50), which becomes undetectable at room temperature. It is
significantly smaller than that reported in previous NMR
studies.\cite{Zalesskii,Zalesskii2,Bush} The magnetic anisotropy energy
derived from the observed anharmonicity is $K_u\sim$10$^{-5}$ eV/spin, it
is comparable to the inhomogeneous exchange energy. The observed
anharmonic parameters bear relevance for understanding the nature of
hybrid excitations,\cite{Sousa} as well as provide input for understanding
the giant effects\cite{Rovillain2} of an electric field on the spin
dynamics of BiFeO$_3$. No magnetic anomalies have been found below the
room temperature. We therefore conclude that the anomalies previously
observed in Raman\cite{Catalan, Cazayous,Singh1} and dielectric
measurements\cite{Redfern} are not related to spin reorientation
transitions.

\section {Acknowledgements}

This work was
supported by the NSF under Grant Nos. DMR-1004568, and DMR-0804109.

\begin{figure}
\includegraphics[scale=.5]{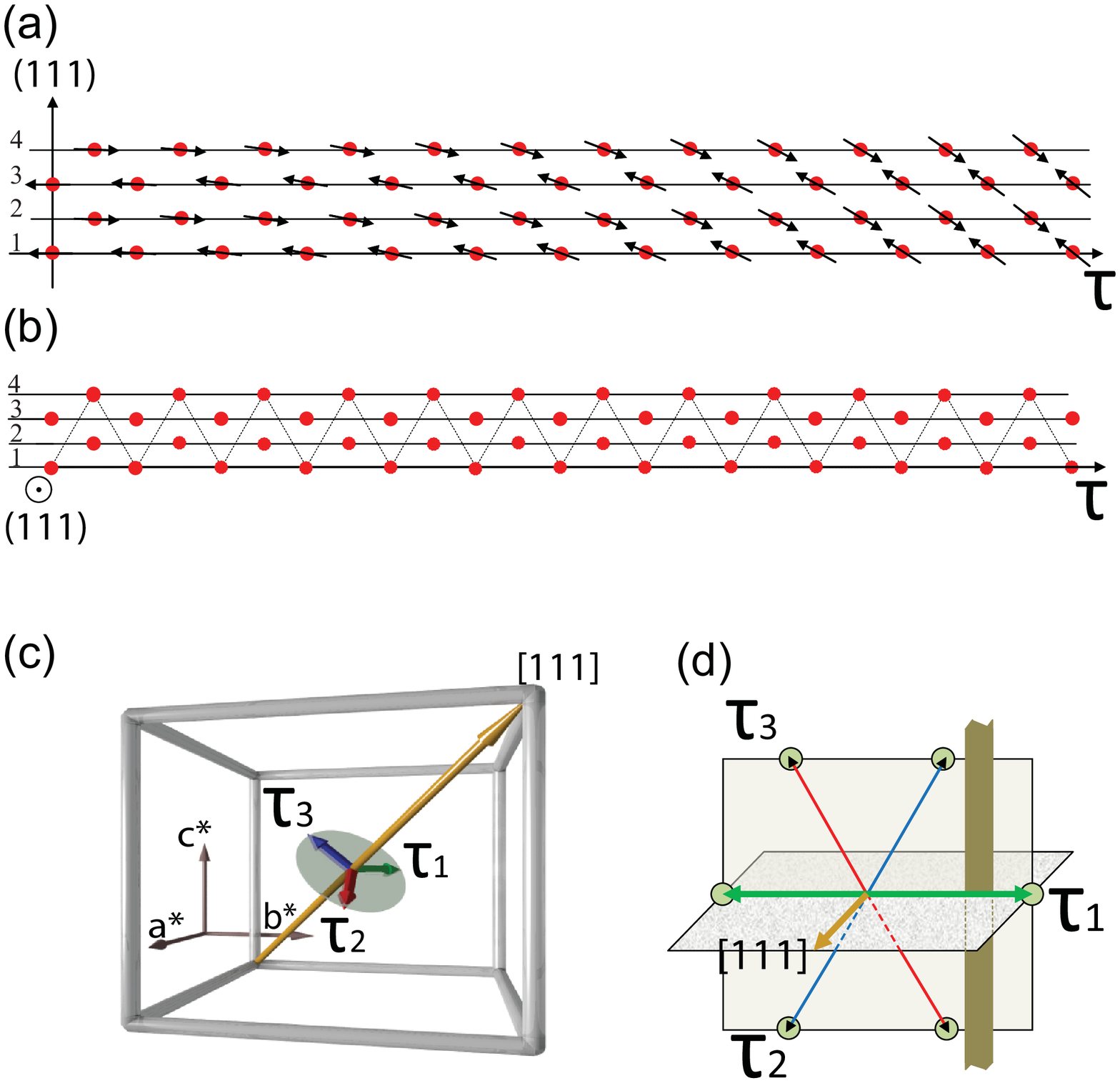}
\caption {\label{povray} (a) Cycloidal magnetic structure of BiFeO$_3$
projected on the plane of the (111) and $\tau$ vectors. About one tenth of
the cycloid period along $\tau$ is shown. (b) View of the same Fe chains,
labeled 1 through 4, projected on the plane normal to (111). (c)
Pseudo-cubic unit cell of BFO and the magnetic modulation wave vectors
$\tau_i$ originating at $Q_0$=(0.5, 0.5, 0.5). (d) Scattering geometry.
The horizontal scattering plane contains $\tau_1$ and (111). Circles
indicate the magnetic peak positions. The experimental resolution function
is sketched with a dark vertical slab.}
\end{figure}

\begin{figure}
\includegraphics[scale=.47]{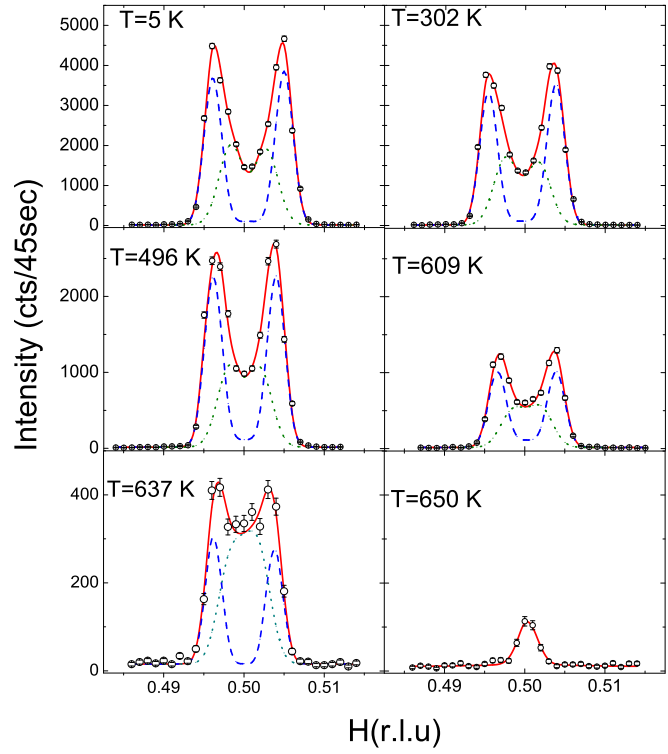}
\caption {\label{profile}Magnetic diffraction peaks in the vicinity of
$Q_0$=(0.5, 0.5, 0.5). The scans are centered at $Q_0$, and the scan
direction is along (H,-H,0.5). Solid lines are results of fits discussed
in the text. Dashed and dotted lines depict the calculated contributions
due to the in-plane and out-of-plane magnetic domains, respectively.}
\end{figure}

\begin{figure}
\includegraphics[scale=.65]{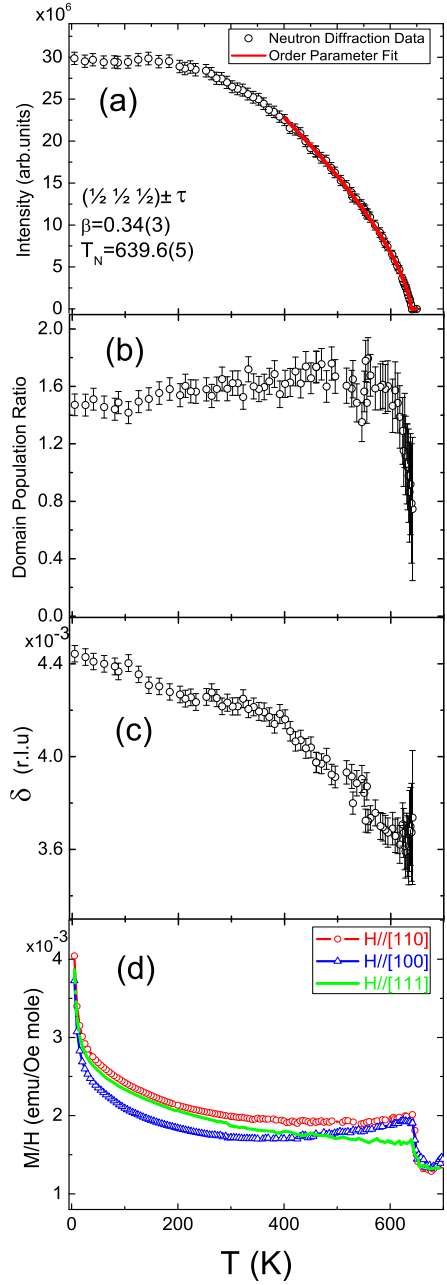}
\caption {\label{parameters}Temperature dependences of (a) Magnetic order
parameter given by the integrated scattering intensity near $Q_0$, (b)
Ratio of the population of the in-plane magnetic domain to the combined
populations of the out-of-plane domains, (c) The magnitude $\delta$ of the
wave vector $\delta$(1,-1,0) of the magnetic modulation, and (d) The
magnetization in the field of $H$=0.2 Tesla for various directions of $H$
(1 emu/Oe = 4$\pi\times$10$^{-6}$ m$^3$).
The solid line in (a) is the result of the power-law fit discussed in the
text.}
\end{figure}

\begin{figure}
\includegraphics[scale=.35]{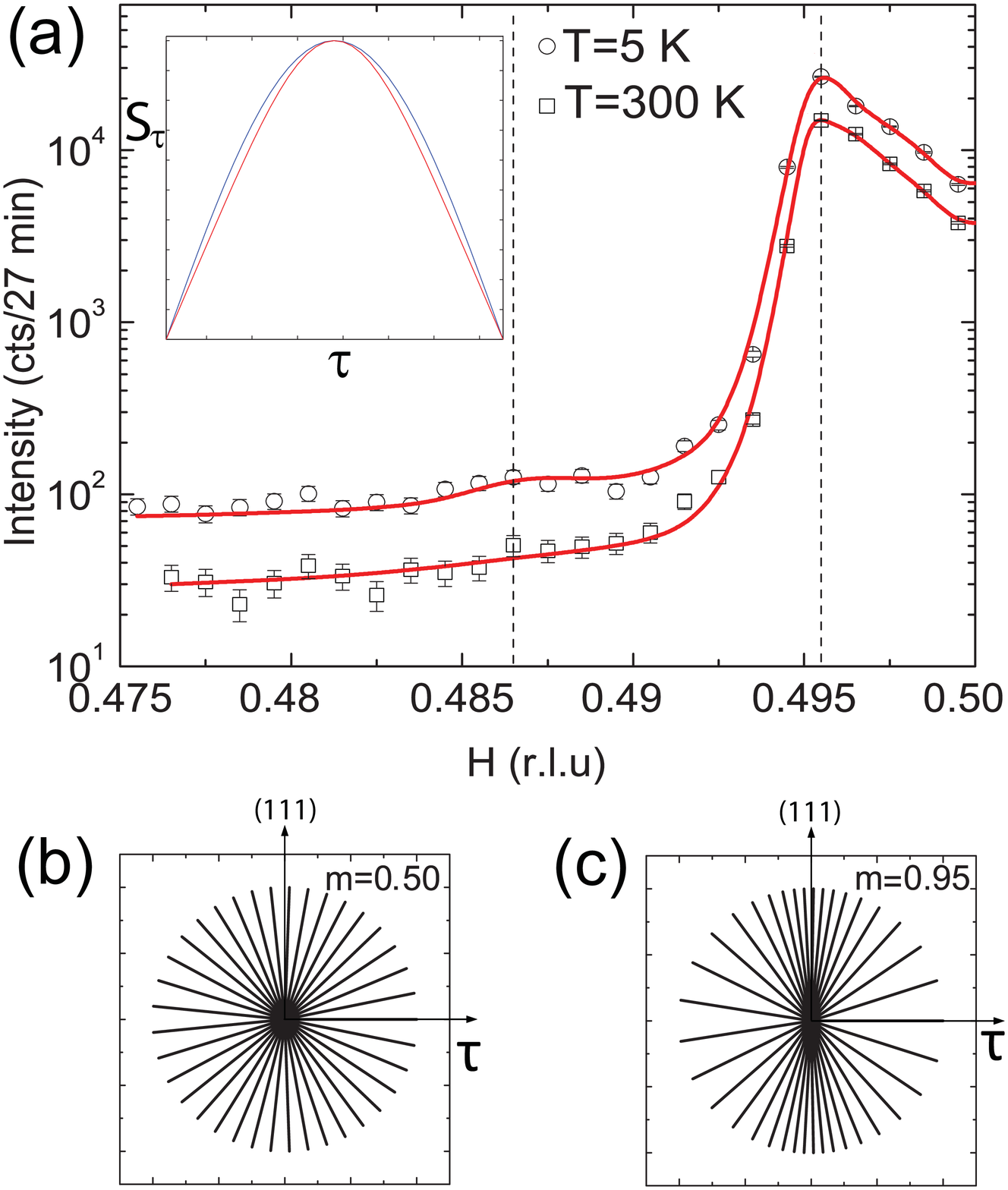}

\caption {\label{weak}(a) Magnetic diffraction peak profiles for $T$=5 K
and 300 K on a logarithmic scale. Solid lines depict results of fits
discussed in the text. Dashed lines indicate positions of the first and
third harmonics. The $T$=300 K scan is shifted down for clarity. The inset
shows the spatial variation of the spin component $S_{\tau}$ along the
modulation propagation direction $\tau$ for the circular cycloid with
$m$=0 (top curve), and for the anharmonic cycloid at $T$=5 K with $m$=0.50
(bottom curve). (b) Fe spins in the anharmonic cycloid with experimentally
determined $m$=0.50 plotted from the same origin. For clarity, only one
third of all spins are shown. (c) The same as (b), but for $m$=0.95. Our
results are inconsistent with the latter value.}

\end{figure}

\end{document}